\documentclass[twocolumn,amsmath,amssymb,superscriptaddress,floatfix]{revtex4-1}
\usepackage[utf8x]{inputenc}
\usepackage{graphicx}
\newcommand*{\mpl}{M_{\rm{Pl}}}
\newcommand*{\lag}{\mathcal{L}}
\newcommand*{\xb}{\tilde{x}}
\newcommand*{\yb}{\tilde{y}}
\newcommand*{\si}{{\rm sign}}

\begin{document}

\title{Cosmological phase space analysis of the $F(X)-V(\phi)$ scalar field and bouncing solutions}

\author{Josue De-Santiago}
\email{josue@ciencias.unam.mx}
\affiliation{Universidad Nacional Aut\'onoma de M\'exico, 04510, D. F., M\'exico}
\affiliation{Institute of Cosmology $\&$ Gravitation, University of Portsmouth, Dennis Sciama Building, Portsmouth, PO1 3FX, United Kingdom}
\affiliation{Depto. de F\'{\i}sica, Instituto Nacional de Investigaciones Nucleares, M\'{e}xico.}

\author{Jorge L. Cervantes-Cota}
\email{jorge.cervantes@inin.gob.mx}
\affiliation{Depto. de F\'{\i}sica, Instituto Nacional de Investigaciones Nucleares, M\'{e}xico.}

\author{David Wands}
\email{david.wands@port.ac.uk}
\affiliation{Institute of Cosmology $\&$ Gravitation, University of Portsmouth, Dennis Sciama Building, Portsmouth, PO1 3FX, United Kingdom}

\date{\today}

\begin{abstract}
  We analyze the dynamical system defined by a universe filled
  with a barotropic fluid plus a scalar field with modified kinetic term of the
  form $\lag=F(X)-V(\phi)$.
  After a suitable choice of variables that allows us to study the phase space of
  the system we obtain the critical points and their stability. We find that they
  reduce to the ones defined for the canonical case when $F(X)=X$. We also study 
  the field energy conditions to have a nonsingular bounce.
\end{abstract}

\pacs{98.80.Cq,95.36.+x}

\maketitle

\section{Introduction}

Scalar fields play an important role in cosmological models because, due
to their simplicity and adaptability, they can account for different interesting
phenomena. They are some of the most popular choices for modeling 
cosmological scenarios such as inflation \cite{Liddle:2000cg}
and dark energy \cite{Copeland:2006wr}, and they also have
been studied in the context of dark matter models \cite{Magana:2012ph},
bounce cosmology \cite{Brandenberger:2012um} and
different unification models of those phenomena \cite{Bertacca:2010ct}.

The proposal that the Lagrangian could be a general function of the kinetic
term $X=-g^{\mu\nu}\partial_\mu \phi \partial_\nu \phi/2$ and the field $\phi$
was introduced in cosmology first in
\cite{ArmendarizPicon:1999rj} in the context of
inflation and then used for dark energy models in \cite{Chiba:1999ka}.
Different particular forms of the
Lagrangian $\lag = p(X,\phi)$ has been studied for different reasons \cite{Copeland:2006wr}.

In this paper we study the general class of models
with sum-separable Lagrangian $\lag = F(X) - V(\phi)$.
Several aspects of this type of scalar fields
have been studied in the literature.
Phenomenology of the inflation models arising from them
\cite{Mukhanov:2005bu,Panotopoulos:2007ky},
topological defects \cite{Babichev:2006cy}, supersymmetry extension
\cite{Adam:2010wu}, boson stars \cite{Adam:2009jp}, unification
models of dark matter and dark energy 
\cite{Bertacca:2010ct,Sharif:2012cu} and unification of dark energy, dark matter and
inflation \cite{Bose:2008ew,DeSantiago:2011qb}.
This model also offers the possibility of being understood as a vacuum
energy density $V$ coupled with a barotropic fluid \cite{Wands:2012vg}.
It reduces to the canonical scalar field when $F(X)=X$.

In section \ref{section_autonomous}
we study the system of autonomous differential equations related to this class
of scalar fields. This method is equivalent to the one used for canonical
scalar fields in \cite{Copeland:1997et} and allows us to identify the
general behavior of the cosmological solutions associated with the present Lagrangian.
This method has been applied to a wide range of cosmological models,
for example
\cite{Kouwn:2012qw,Xu:2012jf,Haghani:2012zq,Escobar:2012cq,Bonanno:2011yx,Boehmer:2011tp,UrenaLopez:2011ur,CervantesCota:2010cb,Leon:2010pu}.
It can be used to determine the presence and stability of solutions
of cosmological interest, such as those with
de Sitter phases or with scaling behaviors.

One possible application for this type of Lagrangian is the generation in
the early Universe of a nonsingular bounce,
in which the state of the Universe goes from collapsing to expanding for $a\ne0$.
The bouncing models have been proposed as alternatives to
inflation \cite{Wands:2008tv,Peter:2008qz} and as a way to evade a singular big-bang, as in
the pre-big-bang scenario \cite{Gasperini:2002bn},
in the ekpyrotic universe \cite{Khoury:2001wf,Buchbinder:2007ad,Lehners:2008vx,Fonseca:2011qi}, or
in other multifield models \cite{Allen:2004vz}.

In section \ref{section_bounce} we study how, when the
density of this field is allowed to be negative, it can drive a bounce.
As the variables defined for previous the dynamical system analysis
are not suitable to study this phenomenon, we redefine the system as in Ref. \cite{Allen:2004vz}
in order to study this case and obtain the conditions to accomplish
a bouncing behavior. To obtain the bounce,
the scalar field has to violate the Null Energy Condition (NEC) possibly
giving rise to instabilities. Some works have been made trying to erase these
instabilities with ghost condensate scalar fields
\cite{ArkaniHamed:2003uy,Creminelli:2006xe,Buchbinder:2007ad} (however see
Ref. \cite{Kallosh:2007ad}). Here we will only consider the dynamics of homogeneous
cosmologies.

\section{Autonomous System for $\lag=F(X)-V(\phi)$ \label{section_autonomous}}

For a spatially flat Friedmann-Lemaitre-Robertson-Walker (FLRW) cosmology
filled with a scalar field with a Lagrangian of the
form $\lag=F(X)-V(\phi)$, where
\begin{equation}\label{kinterm}
  X=-\frac{1}{2}\partial_\mu \phi \partial^\mu \phi \,,
\end{equation}
and a matter component with density $\rho_m$
and equation of state $p_m = \omega_m \rho_m$, the equations of motion are
\begin{equation}\label{fried1}
   H^2=\frac{1}{3\mpl^2} [2XF_X -F + V + \rho_m] \,,
\end{equation}
\begin{equation}\label{fried2}
    \dot{H}= - \frac{1}{2\mpl^2} [2XF_X  + (1+\omega_m) \rho_m] \,,
\end{equation}
where $H$ is the Hubble factor.
These equations can be combined to imply the conservation of
the total energy momentum tensor. We can
suppose additionally the conservation of the scalar field and
barotropic fluid energy momentum tensors
separately, which is the case when there is no 
interchange of energy between the two
components. In that case the barotropic component
satisfies the equation
$\rho_m \propto a^{-3(1+\omega_m)}$ for a
constant $\omega_m$, where $a$ is the scale factor, and the scalar field satisfies 
\begin{equation}\label{cont}
   \frac{d}{dN} (2XF_X-F+V) + 6XF_X = 0,
\end{equation}
where the subindex $X$ means differentiation with respect to that
variable. The time differentiation here has been changed to
$dN=d\log{a}$, a variable that for an expanding FLRW model can be used
as the independent variable instead of the cosmological time, with the
relation $dN=Hdt$.

In order to obtain the autonomous system we define the variables
\begin{eqnarray}
  x &=& \frac{\sqrt{2XF_X-F}}{\sqrt{3}\mpl H} \,, \nonumber \\
  y &=& \frac{\sqrt{V}}{\sqrt{3}\mpl H} \,,
  \label{defs}
\end{eqnarray}
where $x^2$ is proportional to the kinetic part of the energy density
\begin{equation}
  \rho_k = 2XF_X - F \,,
  \label{rhok}
\end{equation}
and $y^2$ to the potential part of the energy density $\rho_V=V$. They are
equivalent to the ones used in the analysis
for canonical scalar fields \cite{Copeland:1997et}.
We will also need to define the auxiliary variables
\begin{eqnarray}
  \label{sigmadef} \sigma &=& -\frac{\mpl V_\phi}{V}\sqrt{\frac{2X}{3|2XF_X-F|}}\si (\dot{\phi}), \\
  \omega_k &= & \frac{F}{2XF_X-F},
\end{eqnarray}
where the former corresponds to the change in time of the potential, as can
be seen if we write it as 
\begin{equation}
  \sigma = -\frac{\mpl}{\sqrt{3|\rho_k|}} \frac{d \log V}{dt},
\end{equation}
and the latter corresponds to the equation of state for the kinetic part of the
Lagrangian, as the kinetic part of the pressure is $P_k=F$.
In the case of a canonical scalar field
$F(X)=X$ and the auxiliary variables turn out to be
$\omega_k=1$ and $\sigma=\sqrt{2/3} \lambda$ for $V\propto e^{-\lambda \phi/\mpl}$ as defined
in Ref. \cite{Copeland:1997et}.

The equation of state of the scalar field can be obtained in terms of
the new variables as
\begin{equation}
  \omega_\phi = \frac{p_\phi}{\rho_\phi} =
  \frac{\omega_k x^2-y^2}{x^2+y^2}.
\end{equation}

The evolution equations for the first two variables ($x$ and $y$) can be written as
\begin{eqnarray}
   \frac{dx}{dN} &=&
   \frac{2XF_{XX}+F_X}{2\sqrt{3(2XF_X-F)}\mpl H}\frac{dX}{dN}
   - x \frac{\dot{H}}{H^2} \,, \label{dxdN1} \\
   \frac{dy}{dN} &=&
   \frac{V_\phi \dot{\phi}}{2\sqrt{3V} \mpl H^2} 
   - y \frac{\dot{H}}{H^2} \,. \label{dydN1}
\end{eqnarray}
The common $\dot{H}/H^2$ factor can be obtained from equation (\ref{fried2})
dividing by $H^2$ and replacing the original for the new variables
\begin{equation}
   \frac{\dot{H}}{H^2} = - \frac{3}{2} \left[
   (1+\omega_m) (1 - y^2 ) + x^2 (\omega_k - \omega_m) \right] \,,
\end{equation}
where we made use of the equation (\ref{fried1}) in the new variables
\begin{equation}
  \label{friedn}
   x^2 + y^2 + \Omega_m = 1.
\end{equation}

Now in order to calculate the first term in the evolution
equation  (\ref{dydN1}) we only have to substitute the values of the new variables
\begin{equation}
   \frac{V_\phi \dot{\phi}}{2\sqrt{3V} \mpl H^2} =
   - \frac{3}{2} \sigma xy \,.
\end{equation}
For the first term in equation (\ref{dxdN1}) we use the continuity
equation (\ref{cont}) that can be written as
\begin{equation}\label{XN}
  \frac{dX}{dN} = - \frac{3F}{(2XF_{XX}+F_X)\omega_k}
  \left( \omega_k+1 - \frac{\sigma y^2}{x} \right) \,,
\end{equation}
so that the evolution equations (\ref{dxdN1}, \ref{dydN1}) become, in terms of
the new variables,
\begin{eqnarray}
   \label {dxdN} \frac{dx}{dN} &=&
   \frac{3}{2}[ \sigma y^2- x (\omega_k + 1) ] +
   \\ & & \frac{3}{2} x \left[ (1+\omega_m) (1 - y^2 ) +
   x^2 (\omega_k - \omega_m) \right] \,,
   \nonumber \\
  \frac{dy}{dN} &=&
  - \frac{3}{2}\sigma yx +
   \label{dydN} \\ \nonumber
   & & \frac{3}{2} y \left[ (1+\omega_m) (1 - y^2 ) + 
   x^2 (\omega_k - \omega_m) \right] \, .
\end{eqnarray}

The evolution equations for the variables $\omega_k$ and
$\sigma$ can be obtained using the equation (\ref{XN}) and the
definition of $X$. But we have to define new auxiliary variables that depend on the
second-order derivatives of the Lagrangian potentials. The evolution equations are
\begin{equation}\label{domegadn}
  \frac{d\omega_k}{dN} = 3 \frac{2\Xi \omega_k + \omega_k -1}{2\Xi +1}
  \left( \omega_k+1 - \frac{\sigma y^2}{x} \right),
\end{equation}
\begin{eqnarray}
  \frac{d\sigma}{dN} &=&  -3\sigma^2 x (\Gamma -1) +
  \label{DSDN}
  \\ \nonumber & & 
  \frac{3\sigma (2\Xi(\omega_k+1) + \omega_k -1)}{2(2\Xi+1)(\omega_k+1)}
  \left( \omega_k+1 - \frac{\sigma y^2}{x} \right),
\end{eqnarray}
where the auxiliary variables are defined as
\begin{eqnarray}
  \Xi &=& \frac{XF_{XX}}{F_X}, \\
  \Gamma &=& \frac{VV_{\phi \phi}}{V^2_\phi}.
\end{eqnarray}

The new second-order derivative variables $\Gamma$, $\Xi$ will have evolution equations in terms of the
dynamical variables and new third-order derivative variables, and so on.
In order to truncate this succession of equations we can consider fixing the functions $F(X)$ and
$V(\phi)$.

The first of these assumptions is to choose the potential related variable $\Gamma$ as a constant. For
it to happen we need
\begin{equation}
  V(\phi) = V_0 (\phi-\phi_0)^{1/(1-\Gamma)}
\end{equation}
for $\Gamma \ne 1$ or
\begin{equation}
  V(\phi) = V_0 e^{-\lambda \phi/\mpl} \label{canonicalG}
\end{equation}
for $\Gamma = 1$.
The second assumption is to consider the case in which
\begin{equation}
  F(X)=AX^\eta,
\end{equation}
where $A$ and $\eta$ are constants, in this case
 $\omega_k=1/(2\eta-1)$
and equation (\ref{domegadn}) is trivially satisfied. In the following we will use
these assumptions.

The dynamical system will be reduced
to Eq. (\ref{dxdN}) for the evolution of $x$,
Eq. (\ref{dydN}) for the evolution of $y$, and an equation
for the evolution of $\sigma$ that due to the choice of $F$
as a power law becomes 
\begin{equation}
  \frac{d\sigma}{dN} = -3 \sigma^2 x (\Gamma -1) +
  \frac{3 \sigma(1-\omega_k)}{2(1+\omega_k)}
  \left( \omega_k+1 - \frac{\sigma y^2}{x} \right).
  \label{dsigmadN}
\end{equation}

\subsection{Critical points}

The autonomous system of equations written above can be analyzed if we consider
its critical points,
in which the equations (\ref{dxdN}) and (\ref{dydN}) are equal to zero, corresponding
to $x$ and $y$ constant. In the first instance we will not consider
the evolution equation (\ref{dsigmadN}),
but if the critical points $(x_0,y_0)$ depend on $\sigma$ they will not
be truly fixed unless we ensure that $\sigma$ is constant.

The variables $x^2$ and $y^2$ correspond to the fraction
of the energy density contained in kinetic and potential energy of
the scalar field, as can be seen from (\ref{friedn}). The condition of
constancy for the critical points implies that these variables have
a constant contribution to the total energy density, which can happen
in three scenarios: (i) if $x^2+y^2$ is equal to one, meaning that all the energy
density comes from the scalar field, (ii) if they are zero, meaning no
contribution, or (iii) if they are between zero and one, corresponding to what is also
known as a scaling solution, meaning that the energy density of the field
scales at the same rate as that of matter. The three behaviors are of cosmological
interest and are present for
the critical points of generally defined parameters $x$ and $y$ if they satisfy
equation (\ref{friedn}).

To present the critical points, we have labeled with
Latin letters those that reduce
in the canonical case to the ones studied in \cite{Gumjudpai:2005ry},
and with Greek letters to the ones with no correspondence.

For $x_a=0$ and $y_a=0$ this corresponds to the scalar field not contributing
to the energy density of the universe.

If $x=0$ and $y\ne 0$ the equation of state becomes
$\omega_\phi = -1$ that is an interesting case from the cosmological point of
view due to the possibility to describe dark energy or inflation phenomena. 
In this case the evolution equation for $x$ reduces to
\begin{equation}
  \frac{dx}{dN}= \frac{3}{2}\sigma y^2,
\end{equation}
which requires $\sigma=0$, a potential that doesn't change with time.
On the other hand the evolution equation for $y$ reduces to
\begin{equation}
  \frac{dy}{dN} = \frac{3}{2} y(1+\omega_m)(1-y^2),
\end{equation}
that can be zero for $\omega_m = -1$ or $y=1$. Both cases correspond to
a FLRW model filled with fluid with equation of state $-1$.
\begin{itemize}
  \item In the first case
$x_\alpha=0$ and $y_\alpha=1$, the Friedmann equation
in the new variables (\ref{friedn}) implies that the matter field has zero
energy density and the only component of the model is the $\phi$ field.

  \item In the second case $x_\beta=0$ and $y_\beta$ is arbitrary, then there are
contributions from the barotropic fluid as well as from the scalar field.
\end{itemize}

If $y=0$ and $x\neq 0$ the potential energy is zero and the equation of state
reduces to $\omega_\phi = \omega_k$. The energy density of the
field is stored in the kinetic part. The evolution equation for $y$ vanishes and the one for $x$ reduces to
\begin{equation}
  \frac{dx}{dN} = \frac{3}{2}x(\omega_k - \omega_m)(x^2-1),
\end{equation}
that can become zero in two cases:
\begin{itemize}
  \item For $x_b=1$, $y_b=0$ this corresponds to the density of the model
    coming entirely from the kinetic part of the field $\phi$.

  \item For $\omega_m = \omega_\phi$, with $y_\gamma=0$ and $x_\gamma$ arbitrary,
the equation of state of the field is the same as the
equation of state of the matter. It corresponds
to a kinetically driven scaling solution. This type of solutions are 
important in cosmology because in the case of
dark energy they have been proposed to alleviate the coincidence
problem \cite{Copeland:2006wr}. This case, however is not completely what in the literature is
called a scaling solution in the sense that it can
only reproduce a constant equation of state of the matter when the
Lagrangian of the field satisfies
\begin{equation}
  \label{cpf}
  F(X)= AX^{(1+\omega_m)/2\omega_m}.
\end{equation}
For example if the energy density of the matter satisfies a relativistic equation of
state, we need $F(X)=AX^2$ such that $\omega_k=1/3$.
The latter happens in the unified dark matter models based in Scherrer's Lagrangian
$F(X) = F_0 + F_m(X-X_0)^2$. It is known \cite{Scherrer:2004au}
that for high energies in which $X\gg X_0$
the model can have a radiation-like behavior and this
is because $F_0$ and $X_0$ can be disregarded, approximating to (\ref{cpf}).
\end{itemize}

The last case is when both $x$ and $y$ are different from zero.
From (\ref{dxdN}, \ref{dydN}) we can see that the critical points satisfy
\begin{equation}
  x= \frac{1}{2\sigma} \left( \omega_k+1 \pm \sqrt{(\omega_k+1)^2-4\sigma^2y^2} \right).
\end{equation}
In this case there are
two different critical points, the first one has the form
\begin{eqnarray}
  x_c &=& \frac{\sigma}{\omega_k+1}, \nonumber\\
  y_c &=&  \frac{\sqrt{(\omega_k+1)^2 - \sigma^2}}{\omega_k+1}. \label{cp1}
\end{eqnarray}
It corresponds to a cosmology filled with the scalar field, as can be seen
from (\ref{friedn}) which in this case corresponds to  $x_c^2+y_c^2=1$
with zero matter density. The equation of state of the system will be
\begin{equation}
  \omega_\phi = \frac{\sigma^2}{1+\omega_k} -1 \,.
\end{equation}

The second nonzero critical point corresponds to
\begin{eqnarray}
  x_d &=& \frac{\omega_m + 1}{\sigma} \,, \nonumber \\
  y_d &=& \frac{\sqrt{(\omega_m+1)(\omega_k - \omega_m)}}{\sigma} \,, \label{cp2}
\end{eqnarray}
where the equation of state in this case is $\omega_\phi=\omega_m$, in other words corresponding
to a scaling solution. In the canonical case with exponential
potential we will recover the scaling
solution of Ref. \cite{Copeland:1997et}.
The fraction of the total energy density
stored in the scalar field will be
\begin{equation}
  x_d^2+y_d^2 = \frac{(1+\omega_k)(1+\omega_m)}{\sigma^2}.
\end{equation}

It is interesting to point that, except for the
canonical case, the Lagrangians studied here with $\omega_k$ and $\Gamma$ constants can't
be reduced to the case $\lag = Xg(Xe^{\lambda \phi})$, that is considered in Ref. \cite{Piazza:2004df}
as the general form for a scalar field with scaling solutions. The difference
from the case studied there is that we are
not considering a coupling between the field and the barotropic fluid as in their case.

The points defined in (\ref{cp1}) and (\ref{cp2}) depend explicitly on $\sigma$, that in general 
is an evolving quantity. It means that, unless the variable $\sigma$ is also
fixed, those points won't be critical points of the system. Setting the evolution equation
for $\sigma$ equal to zero gives us the condition $\Gamma=\Gamma_0$ with
\begin{equation}
  \Gamma_0 = \frac{3+\omega_k}{2(1+\omega_k)}. \label{extracond}
\end{equation}
This relation between the derivatives of potential and kinetic terms in the Lagrangian
has to be accomplished in order to have the critical points (\ref{cp1}, \ref{cp2}). For
the canonical case, as $\omega_k=1$ then the critical points
are fixed only for $\Gamma=1$, that from (\ref{canonicalG}) corresponds to the exponential
potential, as expected. For the noncanonical case as $\omega_k\ne 1$
then there will be a relation between the exponent in the
kinetic and the one in the potential term of the form $\lag = AX^\eta - B (\phi-\phi_0)^n$ with
\begin{equation}\label{extracond2}
  \eta = \frac{n}{2+n} \,,
\end{equation}
if this relation is not satisfied, the critical points won't be truly fixed. In the appendix we show that
when the system satisfies this relation, it is invariant under a set of symmetry transformations, which
 turn allows to reduce the number of degrees of freedom. The same symmetry invariance happens
for the canonical scalar field with exponential potential as proved in Ref. \cite{Holden:1999hm}.

The stability of the critical points can be analyzed by the matrix of the derivatives of
the right hand side of equations (\ref{dxdN}, \ref{dydN}). Analyzing the eigenvalues of the matrix
we obtain the results of table \ref{tab:1}.

The critical point (a) presents a behavior of unstable node for $\omega_k<\omega_m$ which can
drive the scalar field density towards bigger values even if it starts with small density. The
saddle point behavior that was already obtained in the canonical case is recovered here when $\omega_m<\omega_k$.
Point ($\alpha$) corresponds to slow roll behavior as $\sigma=0$ and the potential dominates,
and it can be a saddle point or a stable node depending on the equation of state of the kinetic part.
Point (b) can be stable, unstable or a saddle point. In the canonical case the stable behavior
is not obtained. Points (c) and (d) have the same stability behavior as in the canonical case
except that the conditions get modified by $\omega_k$ as stated in the table. The lines ($\beta$)
and ($\gamma$) are obtained when the equation of state of matter is the same as that of the
kinetic part or the potential part of the Lagrangian and can be stable or unstable.
The cosmological relevance of these solutions is further discussed in the conclusions.

\begingroup
\squeezetable
\begin{table*}
  \caption{Stability and existence of the critical points assuming $-1\le \omega_m \le1$. The 
  points labeled with Latin letters reduce in the canonical case to the ones
  already studied in the literature \cite{Gumjudpai:2005ry}, the points with Greek letters are new.}
  \label{tab:1}
  \begin{ruledtabular}
\begin{tabular}{ccccccc}
    &$x$ & $y$ & Existence & Stability & $\Omega_\phi$ & $\omega_\phi$ \\
    \hline \hline
    (a) & 0 & 0 & Always & Unstable node for $\omega_k<\omega_m$ &0 & - \\
    & &  & &Saddle point for $\omega_m<\omega_k$ & & \\

    \hline 
    ($\alpha$) &0 & 1 & $\sigma=0$ & Saddle point for $\omega_k<-1$ & 1 & $-1$ \\
    & &  & &Stable node for $\omega_k>-1$ & & \\

    \hline
    (b) &1 & 0 & Always & Unstable node for
    $\displaystyle \omega_k>\left\{ \omega_m, \sigma -1  \right\}$& 1&$\omega_k$ \\
    && & &Stable node for
    $\displaystyle\omega_k<\left\{ \omega_m, \sigma -1 \right\}$ & & \\
    && & &Otherwise saddle point & & \\

    \hline
    (c) &$\displaystyle \frac{\sigma}{1+\omega_k}$ &
    $\displaystyle \sqrt{\frac{(1+\omega_k)^2-\sigma^2}{(1+\omega_k)^2}}$&
    $\displaystyle \Gamma=\Gamma_0$ &
    Saddle point for $\displaystyle \sigma^2>(1+\omega_k)(1+\omega_m)$
    & 1 & $\displaystyle \frac{\sigma^2}{1+\omega_k}-1$\\
    && &$\sigma(1+\omega_k)>0$ & Otherwise stable node
    & & \\
    && &$\displaystyle \sigma^2>(1+\omega_k)^2$ & & & \\

    \hline
    (d) &$\displaystyle \frac{1+\omega_m}{\sigma}$&
    $\displaystyle \frac{\sqrt{(\omega_k-\omega_m)(1+\omega_m)}}{\sigma}$&
    $\Gamma=\Gamma_0$ &
    Stable node for $\displaystyle \frac{\sigma^2(8-\omega_k+9\omega_m)}{8(1+\omega_k)(1+\omega_m)^2}<1$& 
    $\displaystyle \frac{(1+\omega_k)(1+\omega_m)}{\sigma^2}$& $\omega_m$\\
    &&&$\omega_m<\omega_k$&Stable spiral otherwise&&\\
    &&&$\sigma>\sqrt{(1+\omega_k)(1+\omega_m)}$&&&\\

    \hline
    ($\beta$) &0 & Arbitrary & $\sigma=0$ and $\omega_m=-1$ & Stable line for $\omega_k>-1$&$y^2$ & $-1$ \\
    && & & Unstable otherwise & & \\

    \hline
    ($\gamma$) &Arbitrary&0&$\omega_k=\omega_m$&Stable line for $x\sigma>1+\omega_k$ & $x^2$&$\omega_m$ \\
    &&&&Unstable otherwise&&\\
  \end{tabular}
\end{ruledtabular}
\end{table*}
\endgroup

\subsection{Critical points at infinity}

The former critical points a, b, c, $\alpha$, $\beta$ and $\gamma$ correspond to the situation in which the
dynamical variables are finite. This is always the case for $x$ and $y$ as Eq. (\ref{friedn}) requires
both variables to be smaller or equal than one. However for $\sigma$ we can see from the definition 
(\ref{sigmadef}) that it can become infinity as $\rho_k$ the kinetic energy density or $V$ the potential tend to zero.
To study this case we make a change of the variable $\sigma$ to $\Sigma=1/\sigma$ and study the
possibility of it becoming zero.

Considering $\omega_k$ and $\Gamma$ constants, the evolution equations (\ref{dxdN}, \ref{dydN}, \ref{dsigmadN}) become
\begin{eqnarray}
   \frac{dx}{dN} &=&
   \frac{3}{2}[ \frac{y^2}{\Sigma} - x (\omega_k + 1) ] +
   \\ & & \frac{3}{2} x \left[ (1+\omega_m) (1 - y^2 ) +
   x^2 (\omega_k - \omega_m) \right] \,,
   \nonumber \\
  \label{dyaux} \frac{dy}{dN} &=&
  - \frac{3xy}{2\Sigma} +
   \\ \nonumber
   & & \frac{3}{2} y \left[ (1+\omega_m) (1 - y^2 ) + 
   x^2 (\omega_k - \omega_m) \right] \, ,\\
  \frac{d\Sigma}{dN} &=&  3 x (\Gamma -1) -
  \frac{3 (1-\omega_k)}{2(1+\omega_k)}
  \left( (\omega_k+1)\Sigma - \frac{ y^2}{x} \right).
\end{eqnarray}
In order to have a critical point at $\Sigma = 0$ it's required that the terms $y^2/\Sigma$ and
$xy/\Sigma$ each vanish. These factors can be computed considering that as $\omega_k$ and $\Gamma$ are constants
the kinetic term is a power-law $F=AX^\eta$, and the potential term is either
a power-law $V=B\phi^n$ or an exponential $V=Ce^{-\lambda \phi/\mpl}$. For the power-law potential the
variable $\Sigma$ has the expression
\begin{equation}
 \Sigma = - \frac{1}{n\mpl} \sqrt{\frac{3(2\eta-1)A}{2}} \si (\dot\phi) \phi X^{(\eta-1)/2} \,,
\end{equation}
and the factors
\begin{eqnarray}
 \frac{y^2}{\Sigma} &=&  - \frac{nB}{3\mpl H^2} \sqrt{\frac{2}{3(2\eta-1)A}}\si (\dot\phi) \phi^{n-1} X^{(1-\eta)/2} \,,\\
 \frac{xy}{\Sigma} &=&  - \frac{n}{3\mpl H^2} \sqrt{\frac{2B}{3}} \si (\dot\phi) \phi^{(n-2)/2} X^{1/2} \,.
\end{eqnarray}

In order to have these two terms equal to zero at the same time as $\Sigma \rightarrow 0$, we requiere $\phi=0$ and $n>2$. With these
conditions the variable $y$ becomes zero too and the evolution equations for $x$ and $y$ get reduced to
\begin{eqnarray}
\frac{dx}{dN} &=& \frac{3}{2}x(x^2-1)(\omega_k - \omega_m) \,, \\
\frac{dy}{dN} &=& 0\,,
\end{eqnarray}
which implies that the critical points occur when $x=0$, $x=1$ or $\omega_k=\omega_m$. These three
cases correspond to the already studied critical points (a), (b) and ($\gamma$), respectively.

The second case happens when the potential is an exponential $V=Be^{-\lambda \phi/\mpl}$.
In this case however $y=\sqrt{V/\rho_c}$ cannot become zero, except for
the trivial case in which $B=0$. This means that
$y^2/\Sigma$ diverges as $\Sigma \rightarrow 0$, which implies that (\ref{dyaux}) cannot be zero and there is
no critical point at infinity.


\subsection{Analysis of the phase space}

\begin{figure}
   \centering
   \includegraphics[width=8cm,keepaspectratio=true]{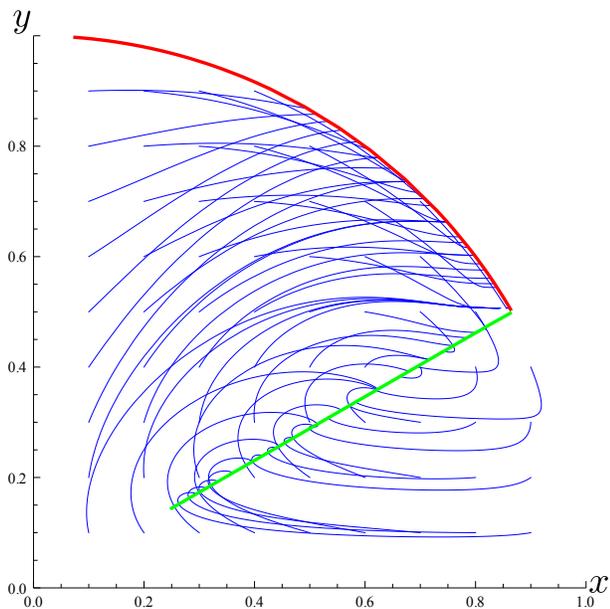}
   \caption{Projection of the phase space along the $(x,y)$ plane for a 
   Lagrangian of the form $\lag= AX^2+B/\phi^{4}$, with initial condition
    $\sigma=1.5$.
   The solutions tend to the critical points (c) and (d) studied in table
   \ref{tab:1}. As $\sigma$ changes, the critical points lie on the
   light green line for the (c) and the red segment of circle for (d).
   This behaviour can be better seen in the
   figures \ref{fig:cuadxs}
   and \ref{fig:cuadys}
   corresponding to different projections of the same system.
   The critical point curves are plotted with the same colors.}
   \label{fig:cuadxy}
\end{figure}

\begin{figure}
   \centering
   \includegraphics[width=8cm,keepaspectratio=true]{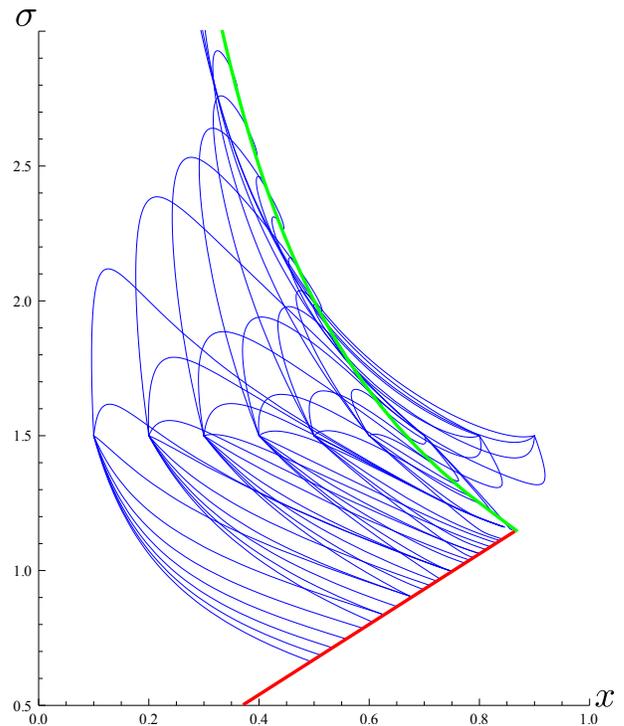}
   \caption{Projection of the phase space along the $(x,\sigma)$ plane for a Lagrangian of
   the form $\lag= AX^2+B/\phi^{4}$. The system is the same as in figures
   \ref{fig:cuadxy} and \ref{fig:cuadys}. The solutions where chosen to
   start in $\sigma=1.5$
   and for a constant $x$ they evolve in different directions due to the different
   values in $y$.
   See the explanation in Fig. \ref{fig:cuadxy}.}
   \label{fig:cuadxs}
\end{figure}

\begin{figure}
   \centering
   \includegraphics[width=8cm,keepaspectratio=true]{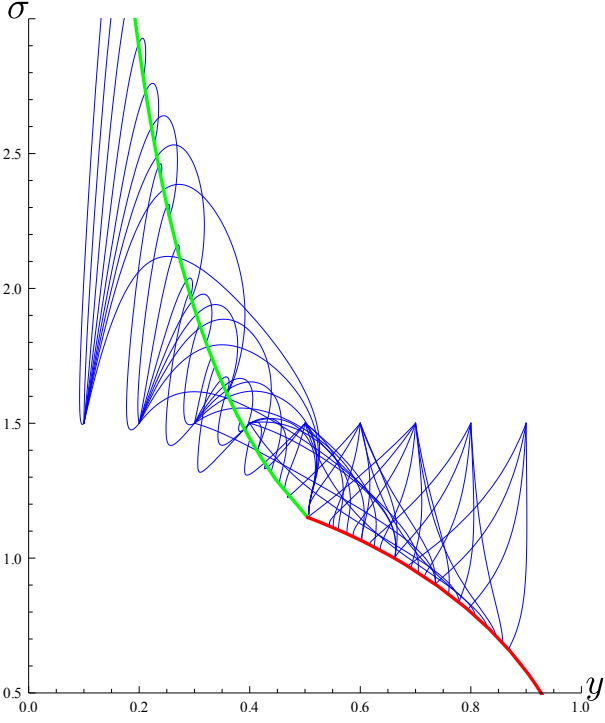}
   \caption{Projection of the phase space along the $(x,\sigma)$ plane for a Lagrangian of
   the form $\lag= AX^2+B/\phi^{4}$. The system is the same as in figures
   \ref{fig:cuadxy} and \ref{fig:cuadxs}. The solutions where chosen to
   start in $\sigma=1.5$
   and they evolve towards different directions depending on the values of $x$ and
   $y$.
   See the explanation in Fig. \ref{fig:cuadxy}.}
   \label{fig:cuadys}
\end{figure}

In this subsection we plot the phase space defined by the equations
(\ref{dxdN}), (\ref{dydN}), (\ref{dsigmadN}) for several potentials.
An important case happens when the potential satisfies equation (\ref{extracond}). For
example this occurs for a Lagrangian of the form $\lag= AX^2+B/\phi^{4}$. In this case the 
extra parameters have values of
$\omega_k=1/3$ and $\Gamma=5/4$. In Figs. \ref{fig:cuadxy}, \ref{fig:cuadxs}, and \ref{fig:cuadys}
we plotted the two dimensional projections of this system for $\omega_m=0$ for an initial condition
of $\sigma=1.5$. We can see how
the solutions approach to the critical points (c) and (d) from table \ref{tab:1} depending on the
values of $\sigma$. The critical points lie on a curve in the 3-dimensional phase space and due to
the evolution of $\sigma$ the solutions tend to different points in those curves.

\begin{figure}
   \centering
   \includegraphics[width=8cm,keepaspectratio=true]{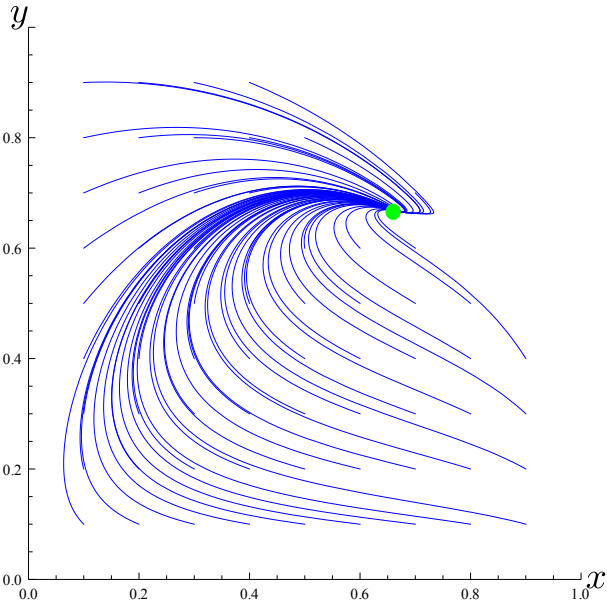}
   \caption{Phase space for the canonical scalar field with exponential potential
   and $\sigma=1.5$. The phase space for this system is two dimensional, unlike in the case
   of nonexponential potentials or noncanonical kinetic terms with three dimensional
   phase spaces. The solutions tend to the critical point (d), scaling solution.}
   \label{fig:canxy}
\end{figure}

As we have stated, for the canonical scalar field the condition
(\ref{extracond}) implies
an exponential potential. In that case $\sigma$ is a constant determined
by the exponent in the potential, as
$V\propto e^{-\sqrt{3}\sigma \phi/(\sqrt{2}\mpl)}$. The phase space in that case is
effectively two dimensional. In Fig. \ref{fig:canxy}
we can see the behaviour for $\sigma=1.5$ with $\omega_m=0$. In this case the solutions approach
only to one point, as there is not evolution in $\sigma$.


\begin{figure}
   \centering
   \includegraphics[width=8cm,keepaspectratio=true]{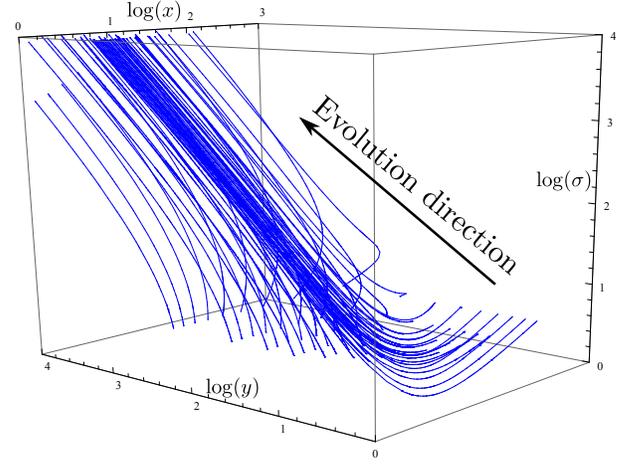}
   \caption{Phase space for the Lagrangian $\lag=AX^2+B\phi^2$. The
   solutions don't tend to any critical point. They evolve towards high
   values of $\sigma$ because this parameter is proportional to $\phi^{-1}$
   for this Lagrangian.}
   \label{fig:cuadcuad3D}
\end{figure}

When the condition (\ref{extracond}) is not satisfied,
we don't have the critical points (c) and (d),
but we can still plot the system. For example, for 
the Lagrangian $\lag=AX^2+B\phi^2$ in which case the values of the
auxiliary parameters are $\omega_k=1/3$ and
$\Gamma=1/2$. In Fig. \ref{fig:cuadcuad3D} we plotted the three dimensional
phase space system for solutions that start with $\sigma=1.5$.
We can see that the system evolves towards big values
of $\sigma$, this happens because $\sigma \propto 1/\phi$ in this case,
and the system goes towards small values of $\phi$. In fact, it crosses
$\phi=0$ in a finite time, in which the variable $\sigma$ is not useful to
describe the system. We can see that the solutions don't tend to any
critical point.

\section{Bounce cosmology \label{section_bounce}}

In this section we consider a nonsingular bounce ($a \ne 0$)
in a FLRW cosmology filled with the scalar field $\mathcal{L}=F(X)-V(\phi)$ and
a barotropic fluid with constant equation of state $\omega_m$.
For a bounce to happen we need the evolution of the scale factor
to go from decreasing to increasing as a function of time.
In terms of the derivative of the scale factor this implies that at the bounce it has to satisfy
$\dot{a}(t_b)=0$ and $\ddot{a}(t_b)>0$. The first condition
can be translated in terms of the Hubble parameter as $H_b=0$,
but this means that the dynamical variables
defined in equations (\ref{defs}) in the last section will diverge.
Besides that, the independent variable $N$ that we have used to parametrize
the evolution of the system is no longer well defined at the bounce,
as $d/dN=H^{-1}d/dt$. Those complications
arise from the fact that our choice of dynamical variables was adjusted to study
a cosmology with increasing $a$. Accordingly, in order to study a
bouncing FLRW metric we need to define new variables adapted to the current
problem.

Also we have to note that the total energy density of the model is zero
at the bounce, which can be seen from the Friedmann equation
(\ref{fried1}). If we suppose that the energy density
of the barotropic component is positive $\rho \propto a^{-3(1+\omega_m)}$
then the energy density of the field has to be negative,
something that will be considered in the definition of the dynamical variables
below.

Now let us define the new set of variables
\begin{eqnarray}
  \xb &=& \frac{\sqrt{3}\mpl H}{\sqrt{|\rho_k|}} \,, \nonumber \\
  \yb &=& \sqrt{\left|\frac{V}{\rho_k}\right|}\si (V) \,,
  \label{xyb}
\end{eqnarray}
where $\rho_k$ is given in Eq.~(\ref{rhok}) and
the absolute values come from the fact that we are interested in the
behavior of both, positive and negative energy densities.
The new independent variable defined in analogy to $N$ is
\begin{equation}
  d\tilde{N} = \sqrt{\frac{|\rho_k|}{3\mpl^2}} dt.
  \label{Nb}
\end{equation}
The evolution equations for the above variables can be then written as
\begin{eqnarray}
  \nonumber \frac{d\xb}{d\tilde{N}} &=&
  -\frac{3}{2} [ ( \omega_k - \omega_m ) \si (\rho_k)
  + (1+\omega_m)(\xb^2- \yb |\yb | ) ] \\
  \nonumber & & + \frac{3}{2} \xb \left[ (\omega_k+1) \xb -
  \sigma \yb |\yb | \si(\rho_k ) \right], \\
  \frac{d\yb}{d\tilde{N}} &=&
  \frac{3}{2} \yb \left[- \sigma
   + (\omega_k+1) \xb -
  \sigma \yb| \yb | \si(\rho_k) \right].
  \label{dydnb}
\end{eqnarray}
In general we also need to evolve $\sigma$, its evolution equation can be obtained
from equation (\ref{DSDN}), transforming to the new variables as
\begin{eqnarray}
  \frac{d\sigma}{d\tilde{N}} &=&  -3\sigma^2 (\Gamma -1) +
  \\ \nonumber & & 
  \frac{3\sigma (2\Xi(\omega_k+1) + \omega_k -1)}{2(2\Xi+1)(\omega_k+1)}
  \left( (\omega_k+1)\xb - \sigma \yb^2 \right),
  \label{dsdnb}
\end{eqnarray}
The above variables are well behaved only for $\rho_k \ne 0$,
so neither of the possible cases of purely potential bounce, nor
a change in sign for $\rho_k$ after the bounce will be studied here.

\subsection{Conditions for a bounce}

In the following we work with the phase space defined by the set of
equations (\ref{dydnb}). Due to the relation between the dynamical variables
$(\xb,\yb)$ and $(x,y)$ from
the previous section, the critical points of both systems coincide. It's
easy to check that the points of table \ref{tab:1} are also critical points of
the new system with the transformation $\xb=1/x$ and $\yb=y/x$,
except for those with $x = 0$ in which the new variables
diverge.

\begin{figure}
   \centering
   \includegraphics[width=8cm,keepaspectratio=true]{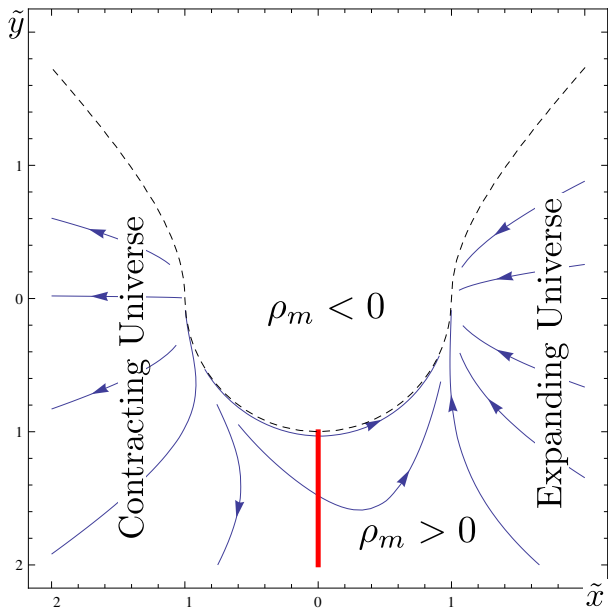}
   \caption{Schematic projection of the phase space for a
 noncanonical nonphantom system with $\rho_k>0$ and $\omega_k=-5$ and $\sigma\sim \sqrt{2/3}$.
   The bounce occurs when the solutions cross the vertical (thick, red) line.}
   \label{fig1}
\end{figure}

\begin{figure}
   \centering
   \includegraphics[width=8cm,keepaspectratio=true]{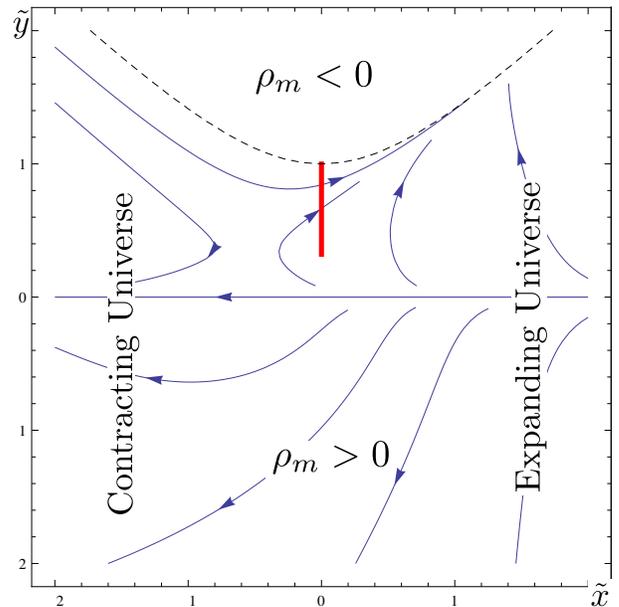}
   \caption{ Schematic projection of the
   phase space for a field with $\rho_k<0$, $\omega_k=1/6$, $\sigma\sim -\sqrt{2/3}$ and $\omega_m=1/3$.
   The bounce occurs when the solutions cross the vertical (thick, red) line.
    There is no purely kinetic ($\rho_V=0$) bounce.}
   \label{fig3}
\end{figure}

\begin{figure}
   \centering
   \includegraphics[width=8cm,keepaspectratio=true]{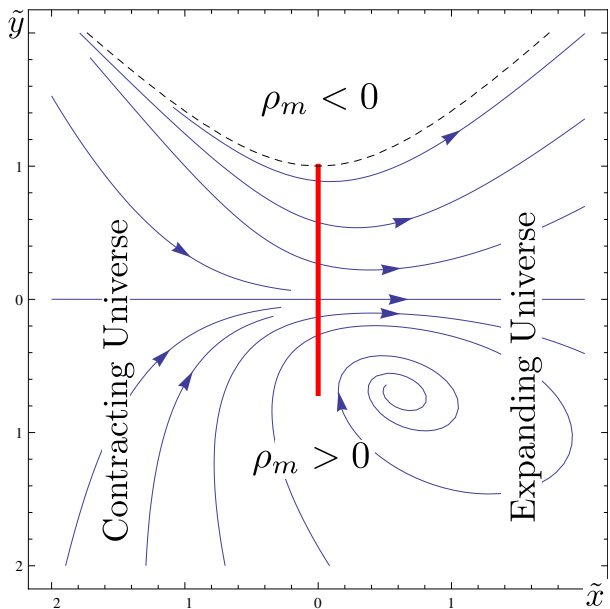}
   \caption{Phase space $(\xb,\yb)$ for the special case of a
   phantom system $F(X)=-X$ and $V\propto e^{-\lambda \phi/\mpl}$ (such that $\sigma$ is
   constant)  plus a barotropic radiation
   component $\omega_m=1/3$. The bounce occurs when the solutions cross the vertical (thick, red) line.
   The spiral in the graph corresponds to the critical point (d) studied in the previous section.}
   \label{fig2}
\end{figure}

\begin{figure}
   \centering
   \includegraphics[width=8cm,keepaspectratio=true]{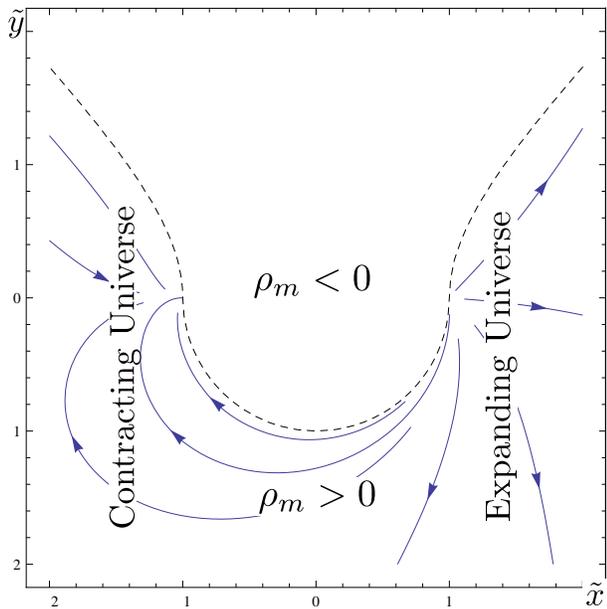}
   \caption{Phase space for the special case of a canonical scalar field with potential
   $V\propto e^{-\lambda\phi/\mpl}$ (such that $\sigma$ is constant) plus a barotropic radiation
   component $\omega_m=1/3$. All the solutions that cross $\xb=0$ have negative $d\xb/d\tilde{N}$ which
   corresponds to recollapse. The bounce is not possible.}
   \label{fig0}
\end{figure}

In this subsection we will use equations (\ref{dydnb}) to study the
evolution of the systems near the bounce. In general we have to consider
also equation (\ref{dsdnb}) to make a representation of the three dimensional
phase space as in the Fig. \ref{fig:cuadcuad3D} of the previous section,
however we will not consider this equation because
we are interested in the behaviour only close to the bounce and the
variable $\sigma$ won't evolve much during this short time. For this reason
the plots of the phase spaces of Fig. \ref{fig1} and \ref{fig3} that will be
studied with more detail in this section,
correspond only to schematic representations
of the phase space near the bounce. For figures \ref{fig2} and
\ref{fig0} the representation corresponds to the actual phase space because
for those Lagrangians $\sigma$ is a constant.


Besides equations (\ref{dydnb}) the system
has to satisfy the Friedmann constraint (\ref{fried1}),
which translates into
\begin{equation}
  \xb^2 - \yb |\yb| - \tilde{\Omega}_m = 1 \times \si (\rho_k),
\end{equation}
where $\tilde{\Omega}_m=\rho_m / |\rho_k|$ corresponds to a dimensionless
density parameter for the
barotropic fluid component. As $\tilde{\Omega}_m$ is assumed to be
nonnegative, we obtain the expression
\begin{equation}
  \xb^2 - \yb |\yb| \ge 1 \times \si (\rho_k),
  \label{allowed}
\end{equation}
which defines the allowed regions of the
phase space.
For the nonphantom case $\rho_k>0$, the Friedmann constraint becomes
\begin{equation}
  \xb^2 - \si (V) \yb^2  \ge 1,
\end{equation}
which for $\yb$ positive corresponds to the region inside the branches of
the hyperbola $\xb^2 - \yb^2 = 1$, and for $\yb$ negative to
the region outside the circle defined by $\xb^2+\yb^2=1$, as can be seen in
figure \ref{fig1}.
In the $\rho_k<0$ case the condition (\ref{allowed}) translates into
\begin{equation}
  \xb^2 - \si (V) \yb^2 \ge -1,
\end{equation}
which for $\yb>0$ corresponds to the region below the hyperbola
$\yb^2-\xb^2=1$. For $\yb<0$ this condition
is satisfied for all the values, as we can see in the
figure \ref{fig2}.

From the definitions (\ref{xyb}) we can see that
\begin{itemize}
  \item $\xb > 0$ corresponds to the regime of an expanding cosmology,
  \item $\xb < 0$ corresponds to a contracting cosmology,
  \item $\xb = 0$ corresponds either to a bounce, a recollapse or a static cosmology.
\end{itemize}
For the case of $\xb=0$ we can use the information contained in the derivative to
study whether we are dealing with a bounce or a recollapse:
\begin{itemize}
  \item $\frac{d\xb}{d\tilde{N}} > 0$ corresponds to a bounce,
  \item $\frac{d\xb}{d\tilde{N}} < 0$ corresponds to a recollapse,
  \item $\frac{d\xb}{d\tilde{N}} = 0$ gives not enough information and one has to
    consider higher derivatives or analyze the neighboring phase space.
\end{itemize}

To see which of the above cases occurs in the phase space of our system
we use $\xb=0$ in the evolution equations
(\ref{dydnb}). In particular for the evolution of $\xb$ we obtain
\begin{equation}
  \frac{d\xb}{d\tilde{N}} =
  -\frac{3}{2}\left[ \left( \omega_k - \omega_m \right) \si (\rho_k)
  - (1+\omega_m)  \yb |\yb| \right] .
\end{equation}
As we stated above this expression has to be positive for a bounce,
which implies a condition in the parameter $\yb$ as
\begin{equation}
  \yb > \sqrt{\left| \frac{\omega_k-\omega_m}{1+\omega_m}\right|} \si (\rho_k(\omega_k-\omega_m)) \,.
\label{bounce1}
\end{equation}
In addition we also have the condition (\ref{allowed}) for the case $\xb=0$
\begin{equation}
  \yb \le -1\times \si (\rho_k).
  \label{bounce2}
\end{equation}

To analyze the above conditions we first suppose $\rho_k>0$. In this case
the inequality (\ref{bounce2}) transforms to $\yb \le -1$ and (\ref{bounce1}) to
$\yb^2<(\omega_m-\omega_k)/(1+\omega_m)$. For the two conditions to
be satisfied in an interval of $\yb$ is necessary to have $\omega_k<-1$.
For example, in the case of a canonical scalar field, one has $F(X)=X$
and consequently $\rho_k>0$, but as $\omega_k=1$ the system of a barotropic
component and a canonical scalar field cannot give rise to a bounce, as
was already shown in \cite{MolinaParis:1998tx}. This can be seen in the phase space of figure
\ref{fig0} in which all the solutions that cross the $\yb$ axis move
from positive to negative values of $\xb$, corresponding to recollapse.

For $\rho_k<0$ the conditions for the bounce become
\begin{equation}
  \sqrt{\left| \frac{\omega_k-\omega_m}{1+\omega_m}\right|} \si (\omega_m-\omega_k)
  < \yb \le 1 \,,
\end{equation}
which can be satisfied for an interval of $\yb$ as long as $\omega_k>-1$. The
original phantom field with $F(X)=-X$ satisfies $\omega_k=1$ and, as shown in 
figure \ref{fig2}, can have a bounce behavior.

In order to have a purely kinetic bounce, in other words one with $\yb=0$ the conditions
above state that the density of the scalar field $\rho_k$ has to
be negative and $\omega_k>\omega_m$. Figure \ref{fig3} shows a case in which
the later is not accomplished and then there is no purely kinetic bounce.

\begin{figure}
   \centering
   \includegraphics[width=8cm,keepaspectratio=true]{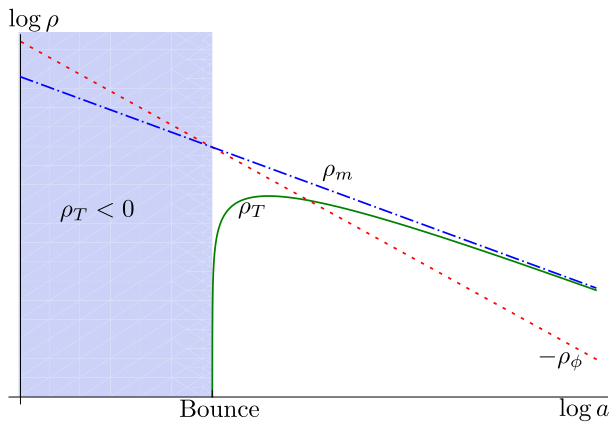}
   \caption{The densities of the barotropic fluid (blue, dash-doted line), the scalar field
   (red, doted line), and the
   total density of the universe (green, continuous line)
   respectively as a function of the scale factor. The total
   energy density tends to zero at the bounce and for smaller values 
   of $a$ is negative, which is forbidden.}
   \label{fig4}
\end{figure}

The two conditions in the previous paragraph can be generalized. First,
to obtain a bounce one needs the total energy density of the field to
be negative in order to compensate for the positive barotropic energy density
in the Friedmann equation 
\begin{equation}
  H^2=\frac{1}{3\mpl^2}(\rho_\phi + \rho_m)=0,
\end{equation}
where $\rho_\phi=\rho_V+\rho_k$ is the total energy density in the field.
Moreover, the total equation of state of the field $\omega_\phi$
has to be bigger than that of the barotropic fluid in order to have
a positive energy density for $a>a_{bounce}$, as can be seen in 
figure \ref{fig4}. Otherwise we will be
dealing with a system that exhibits positive energy density only
for $a<a_{bounce}$ corresponding to a recollapse.

The above two conditions are in fact the same as those in the expressions
(\ref{bounce1}) and (\ref{bounce2}) in terms of the dynamical variables. For the first one,
the negativity of the energy density $\rho_k+\rho_V$ can be translated
as $1 \times \si (\rho_k) + \rho_v/|\rho_k|<0$ or from the definitions
of the variables (\ref{xyb}) as
\begin{equation}
\yb<-1\times \si (\rho_k),
\label{neg1}
\end{equation}
which corresponds to expression (\ref{bounce2}).
The condition on the total
equation of state of the field, in terms of the dynamical variables can be written
as
\begin{equation}
  \frac{\omega_k-\yb |\yb| \si (\rho_k)}{1+\yb |\yb| \si (\rho_k)} > \omega_m,
\end{equation}
which can be transformed into (\ref{bounce1}) after some algebra and using the expression (\ref{neg1}).

\begin{figure}
   \centering
   \includegraphics[width=8cm,keepaspectratio=true]{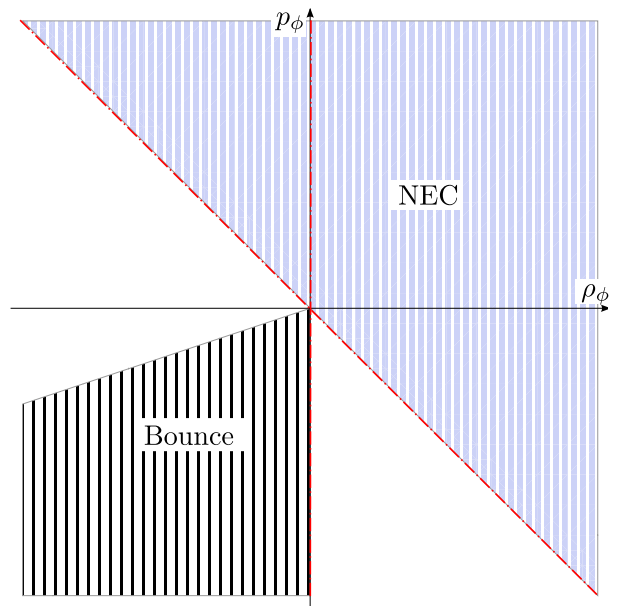}
   \caption{The bottom left region corresponds in the $\rho-p$ plane
   to the part which can drive a bounce, with $\rho<0$ and
   $\omega_\phi>\omega_m$ with $\omega_m=1/3$. The upper right region is
   the one that satisfies the Null Energy Condition. The dash-doted lines
   cannot be crossed by k-essence Lagrangians like the ones consedered here \cite{Vikman:2004dc}.}
   \label{fig5}
\end{figure}

The conditions on the field to have a negative energy density and an
equation of state greater than that of matter implies a violation of
the Null Energy Condition (NEC) that states that $\rho_\phi + p_\phi$ be positive, as can be seen
in Fig. \ref{fig5}. In the last years an extensive literature has been produced studying
fields that violate the NEC.
The main reason for that interest is because the
current measurements of the dark energy equation of state slightly favor models with
$\omega_{de}<-1$ \cite{Copeland:2006wr}.
However fields violating the NEC might have several types of instabilities, for example imaginary
sound speed which results in an increase of inhomogeneities in small periods of time \cite{Vikman:2004dc,Xia:2007km},
or decay of the vacuum into negative energy particles of the field plus positive
energy particles \cite{Carroll:2003st,Garriga:2012pk,Sawicki:2012pz}. The inclusion of higher order terms in the Lagrangian
has been proposed as a method to obtain particles with positive energy in the
so called Ghost condensate models
\cite{ArkaniHamed:2003uy,Creminelli:2006xe,Buchbinder:2007ad,Creminelli:2007aq},
however usually these extra terms add new stability problems to the models, and is not clear if there is a well behaved
high energy theory to account for them \cite{Adams:2006sv,Adams:2006sv}.
Due to those problems, a recent series of works has been published studying fields in which the
introduction of certain symmetries can ensure the stability of the model in spite of breaking
the NEC \cite{Creminelli:2010ba,Easson:2011zy,Qiu:2011cy}.
However those models, so called Galileons, have dynamics which was not studied in this paper.


\section{Conclusions \label{conclu}}

As we have seen, the system of equations (\ref{fried1}, \ref{fried2})
for the Lagrangian $\lag = F(X) - V(\phi)$
can be rewritten in terms of the dynamical variables (\ref{defs})
as (\ref{dxdN}, \ref{dydN}). This system
allows us to understand the dynamical behavior of the universe under different
initial conditions. The critical points
and their stability are summarized in table \ref{tab:1}.
This system
is naturally adapted to study Lagrangians with kinetic terms
of the type $F(X) \propto X^\eta$ and potentials
$V(\phi)\propto \phi^{1/(\Gamma-1)}$ or $V(\phi) \propto e^{-\lambda \phi}$
such as those studied for k-inflation in \cite{Panotopoulos:2007ky}.
The canonical case and its critical points are recovered for $F(X)=X$.

In general the critical points ($\alpha$), ($\beta$), ($\gamma$), (c) and (d)
are present only for particular choices of the Lagrangian $\lag =F(X)-V(\phi)$,
as happens for the canonical scalar field in which the points (c) and (d)
are only present for exponential potentials. The conditions for their existence
are summarized in table \ref{tab:1}.

The point ($\alpha$) corresponds to the slow roll scenario in which
the potential dominates ($y=1$) and its derivative is zero ($\sigma=0$).
The case with $\omega_k<-1$ is interesting for inflationary models as it
corresponds to a saddle point, offering an explanation of how the
universe could enter in the slow roll regime and exit eventually. For that case it is also
necessary to study the dynamical behavior of $\sigma$ to understand the conditions
for it to evolve towards zero, something that was not analyzed in this paper.

The potential dominated line ($\beta$) has the cosmologically interesting behavior
of an equation of state of $-1$, however it requires that the barotropic
fluid has the same behavior, something that is very restrictive.

The kinetic dominated line ($\gamma$) corresponds to critical points of the system
only when $\omega_k=\omega_m$, for example if
$F(X) \propto X^2$ when the barotropic fluid is radiation. It can happen for
example in the purely kinetic unified model studied in Ref.~\cite{Scherrer:2004au},
in which the proposed Lagrangian behaves as a radiation fluid for high energies.
An interesting extension to this purely kinetic model is the addition 
of a potential term to the Lagrangian which could leave the kinetic dominated
line stable at early times, setting the initial conditions necessary for a later evolution
as dark matter plus dark energy if the potential becomes flat at late times,
see also Ref. \cite{DeSantiago:2011qb}. 

The scalar field dominated solution
(c) and the scaling solution (d) are not in general critical points of the system except for the
case in which the potentials in the Lagrangian satisfy the particular
relation (\ref{extracond}). This relation for the canonical scalar field means
that the potential has to be exponential, and for the noncanonical field means
that the Lagrangian has to satisfy (\ref{extracond2}).
As in the canonical case, the scaling solution corresponds
to a stable node or a stable spiral, and the scalar field dominated solution
behaves as a stable node or saddle point. If condition (\ref{extracond}) is not
satisfied, even if $dx/dN=0$ and $dy/dN=0$ for a particular time, the variables
will evolve because the time dependence on $\sigma$ will drive
$x'\ne0$ and $y'\ne0$ as time passes.
However in the cases in which (\ref{extracond}) is satisfied we can obtain
scaling solutions despite the fact that the Lagrangian cannot be reduced to
the form $\lag = Xg(Xe^{\lambda \phi})$, which is studied in Ref. \cite{Piazza:2004df}
as the general form of scalar fields with scaling solutions; but
we are not considering here an interaction with the matter component as
in that case.

In order to study a bouncing cosmology we had to redefine the dynamical variables to some
more suited to the problem as (\ref{xyb}). We obtained the  conditions
(\ref{bounce1}, \ref{bounce2}) necessary for a bounce.
In the phase space it is seen as the possibility to have a crossing
of the $\yb$ axis from the negative to the positive $\xb$ region.

The dynamical variables
$(\xb,\yb)$ and $(x,y)$ are related by the transformation
$\xb=1/x$ and $\yb=y/x$ which means that the critical points of both dynamical systems 
coincide when both are valid. This happens when $\rho_k$ and $\rho_V$ 
are positive, otherwise the variables $x$, $y$ are not defined, and when $x\ne0$.
It can be seen that the points of table \ref{tab:1} are also critical points of
the new system except for those with $x = 0$.

We split the analysis of the bouncing system in two cases,
$\rho_k$ negative (phantom scalar field) and $\rho_k$ positive,
and obtained that in order to have a bounce we need $\omega_k>-1$ for the
first case and $\omega_k<-1$ for the second one.

For a canonical scalar field, we know that a negative potential can lead to
a crossing of the $\xb$ axis ($H=0$) only for recollapse, and not for
a bounce. Here we showed that for certain values of $\omega_k$ a bounce is possible
even for $\rho_k$ positive, giving the possibility of a potentially driven bounce.
We also showed that the conditions (\ref{bounce1}, \ref{bounce2}) obtained in terms of the dynamical variables
can be ultimately understood as 
$\rho_\phi<0$ and $\omega_\phi>\omega_m$, better seen from figure \ref{fig4}
as the conditions to have zero energy density at the bounce and positive
energy density immediately after and immediately before it.

We showed that the field has to violate the Null Energy Condition in order
to account for the bounce, as can be seen in figure
\ref{fig5}. This is a well known result that can have implications concerning the
stability of the field.
 It this paper we didn't deal with the inhomogeneous perturbations, however 
it has been argued that this type of Lagrangians have both classical and quantum stability problems
when they violate the NEC \cite{Vikman:2004dc,Kallosh:2007ad}.
All the former arguments make us conclude that possibly fields as simple as F-V are not good candidates
to violate NEC and therefore to produce a bounce. The study of other types of fields might be in order
but it escapes the purpose of the present paper where only the homogeneous dynamics of the fields was
considered.

\acknowledgments
We thank I. Sawicki and A. Vikman for helpful comments. 
JDS is supported by CONACYT Grant 210405 and JLCC by Grant 84133-F.
DW is supported by STFC grant ST/H002774/1.
JDS acknowledges ICG, University of Portsmouth for their
hospitality.

\appendix*
\section{Symmetry for particular Lagrangians}

The critical points (b) and (d) from table \ref{tab:1} exist only for canonical scalar fields
with exponential potential or for
scalar fields whose Lagrangians are of the form 
\begin{equation}\label{powerlaw}
\lag = AX^\eta - B (\phi-\phi_0)^n
\end{equation}
with
\begin{equation}\label{symi}
  \eta = \frac{n}{2+n} \,.
\end{equation}
In these cases the system presents a symmetry
that allows the number of degrees of freedom to be reduced to two, and the
dynamical system to be described only by $x$ and $y$.
For the canonical scalar field with exponential potential this symmetry was described
in \cite{Holden:1999hm}.

The equations of motion (\ref{fried1}-\ref{cont}) plus the continuity equation
for the barotropic component can be written for a Lagrangian of the form
(\ref{powerlaw}) as
\begin{eqnarray}
  \label{fried1_aux}  &&
  H^2 = \frac{1}{3\mpl^2} [ (2\eta-1)A X^\eta + B\phi^n + \rho_m] \,,
   \\ \label{fried2_aux} &&
   H \frac{dH}{dN} = - \frac{1}{2\mpl^2} [2\eta A X^\eta  + (1+\omega_m) \rho_m] \,,
   \\ &&
   \frac{d\rho_m}{dN} = - 3 (1+\omega_m) \rho_m \,,
   \\ \label{cont_aux} &&
   \frac{d}{dN} ( (2\eta-1)AX^\eta + B \phi^n) = -6 \eta A X^\eta \,,
\end{eqnarray}
where for simplicity we considered $\phi_0=0$. Here $\phi$, $X$, and $\rho_m$ are the independent variables and 
the transformation
\begin{eqnarray}
  \phi &\rightarrow & \xi^{2\eta} \phi \,, \nonumber \\
  X &\rightarrow& \xi^{2n} X \,, \nonumber \\
  \rho_m &\rightarrow & \xi^{2n\eta} \rho_m \,,
  \label{transf}
\end{eqnarray}
will leave invariant the equations of motion as long as the Hubble parameter also transforms
as $H \rightarrow \xi^{n \eta}H$, but its transformation is already determined by
the relation
\begin{equation}
  X=\frac{1}{2} \left( H \frac{d\phi}{dN} \right)^2,
  \label{Xdef2}
\end{equation}
which implies that $H$ transforms as $\xi^{n-2\eta}H$.
In order to have the correct transformation relation for the Hubble parameter then
it is needed that $n\eta=n-2\eta$ which is equivalent to the relation
(\ref{symi}), only in that case the transformation (\ref{transf}) will represent a
symmetry of the system leaving invariant the equations of motion.

The presence of the symmetry transformation (\ref{transf}) when (\ref{symi}) holds means that
the number of degrees of freedom in the equations of motion can be reduced by one.
For this a set of variables invariant under the transformation needs to be defined, in
this case $x$ and $y$ are already invariant. Any dynamical variable can be written in
terms of those two variables, for example $\sigma$ satisfies the relation
\begin{equation}
  \sigma = s \left( \frac{x}{y} \right)^{2/n}\,,
\end{equation}
where $s$ is a constant defined by the parameters in the Lagrangian as
\begin{equation}
  s \equiv - \sqrt{\frac{2}{3}}\mpl n B^{1/n} (A(2\eta-1))^{-1/2\eta} \,.
\end{equation}
From this relation, the dynamical system can be rewritten as
\begin{eqnarray}
    \frac{dx}{dN} &=&
   \frac{3}{2}[ sxy\left( \frac{y}{x} \right)^{2/(\omega_k+1)}  - x (\omega_k + 1) ] +
   \\ & & \frac{3}{2} x \left[ (1+\omega_m) (1 - y^2 ) +
   x^2 (\omega_k - \omega_m) \right] \,,
   \nonumber \\
  \frac{dy}{dN} &=&
  - \frac{3}{2} s x^2 \left( \frac{y}{x} \right)^{2/(\omega_k+1)} +
    \\ \nonumber
   & & \frac{3}{2} y \left[ (1+\omega_m) (1 - y^2 ) + 
   x^2 (\omega_k - \omega_m) \right] \, ,
\end{eqnarray}
corresponding to only two equations for two variables.

\bibliography{biblio}{}

\end{document}